\documentclass[12pt]{article}
\usepackage{axodraw}

\catcode`@=12
\begin{document}
\thispagestyle{empty}
\begin{flushright} 
UCRHEP-T335\\ 
June 2002\
\end{flushright}
\vspace{0.5in}
\begin{center}
{\LARGE	\bf New Perspective on Fermion Mass Matrices\\}
\vspace{2.0in}
{\bf Ernest Ma\\}
\vspace{0.2in}
{\sl Physics Department, University of California, Riverside, 
California 92521\\}
\vspace{2.0in}
\end{center}
\begin{abstract}\
It is shown that any $n \times n$ Dirac fermion mass matrix may be written as 
the sum of $n$ states of equal ``mass''.  However, these states are in  
general not orthogonal.  Thus the texture of any such fermion mass matrix may 
be understood as the nonzero overlap among these states.
\end{abstract}
\newpage
\baselineskip 24pt

The fermions of the standard model of particle interactions, i.e.~the quarks 
and leptons, have definite masses and mix with one another.  Consider first a 
$2 \times 2$ quark mass matrix.  In terms of its mass eigenstates, it can 
always be written as
\begin{equation}
-{\cal L}_m = m_1 \bar q_{1L} q_{1R} + m_2 \bar q_{2L} q_{2R} + H.c.
\end{equation}
I now choose to write the above as follows.
\begin{eqnarray}
-{\cal L}_m &=& (\sqrt {m_1} A_{11} \bar q_{1L} + \sqrt {m_2} A_{21} 
\bar q_{2L})(\sqrt m_1 A^*_{11} q_{1R} + \sqrt {m_2} A^*_{21} q_{2R}) 
\nonumber \\ &+& (\sqrt {m_1} A_{12} \bar q_{1L} + \sqrt {m_2} A_{22} 
\bar q_{2L})(\sqrt {m_1} A^*_{12} q_{1R} + \sqrt {m_2} A^*_{22} q_{2R}) 
+ H.c.
\end{eqnarray}
This implies that
\begin{equation}
|A_{11}|^2 + |A_{12}|^2 = |A_{21}|^2 + |A_{22}|^2 = 1, ~~~ A_{11} A^*_{21} 
+ A_{12} A^*_{21} = 0,
\end{equation}
i.e. $A_{ij}$ should be a unitary matrix but otherwise unrestricted. 
In particular, let it be given by
\begin{equation}
A = {1 \over \sqrt 2} \left[ \begin{array} {c@{\quad}c} 1 & 1 \\ 1 & -1 
\end{array} \right].
\end{equation}
Then Eq.~(2) becomes
\begin{equation}
-{\cal L}_m = {1 \over 2} (m_1 + m_2) (\bar \psi_{1L} \psi_{1R} + 
\bar \psi_{2L} \psi_{2R}) + H.c.,
\end{equation}
where
\begin{eqnarray}
\psi_1 &=& {1 \over \sqrt {m_1 + m_2}} (\sqrt {m_1} q_1 + \sqrt {m_2} q_2), \\ 
\psi_2 &=& {1 \over \sqrt {m_1 + m_2}} (\sqrt {m_1} q_1 - \sqrt {m_2} q_2).
\end{eqnarray}
In other words, in the basis of $\psi_{1,2}$, the single ``mass'' term 
$(m_1+m_2)/2$ appears for both states.  Of course, the true masses are 
$m_1$ and $m_2$ of Eq.~(1) which is identical to Eq.~(2) as well as to 
Eq.~(5).  What has been done is just a new way of writing the same thing. 
This is possible because $\psi_1$ is not orthogonal to $\psi_2$ as shown 
clearly by Eqs.~(6) and (7), although they are certainly independent. 
The same information is recovered by noting that the overlap between $\psi_1$ 
and $\psi_2$ is given by $(m_1 - m_2)/(m_1 + m_2)$.

The extension of the above argument to the $n \times n$ case is 
straightforward.  For equal ``mass'' terms, $A_{ij}$ should then be of the form
\begin{equation}
A = {1 \over \sqrt 3} \left[ \begin{array} {c@{\quad}c@{\quad}c} 1 & 1 & 1 \\ 
1 & \omega & \omega^2 \\ 1 & \omega^2 & \omega \end{array} \right]
\end{equation}
in the $3 \times 3$ case, where $\omega^3 = 1$, and
\begin{equation}
A = {1 \over 2} \left[ \begin{array} {c@{\quad}c@{\quad}c@{\quad}c} 1 & 1 & 
1 & 1 \\ 1 & i & -1 & -i \\ 1 & -1 & 1 & -1 \\ 1 & -i & -1 & i \end{array} 
\right]
\end{equation}
in the $4 \times 4$ case.  Using the discrete symmetry $Z_n$, this is easily 
generalized to any $n$.  Note that in the $3 \times 3$ case, the common 
``mass'' is $(m_1 + m_2 + m_3)/3$ and each of the three overlaps between 
pairs of the three corresponding $\psi_i$ is identical as well, i.e.
\begin{equation}
\zeta = {m_1 + \omega m_2 + \omega^2 m_3 \over m_1 + m_2 + m_3}.
\end{equation}
Given the numerical value of the complex parameter $\zeta$ and that of the 
common ``mass'', it is clear that $m_{1,2,3}$ may be extracted.  This 
procedure is applicable for both the $up$ and $down$ quarks as well as 
the charged leptons.

The possibility that $\psi_i$'s have the same ``mass'' suggests some sort of 
universal underlying dynamics.  Fermions of each sector are states 
of equal ``mass'' which are independent but not orthogonal.  The spacetime 
evolution of these states requires them to rearrange in terms of true mass 
eigenstates and the usual description is recovered.  The puzzle of 
hierarchical fermion masses is now seen in a different light.  It is not 
necessary to look for a fundamental mechanism which generates very different 
masses.  It may be such that equal or nearly equal ``masses'' are generated, 
but without the orthogonality of the corresponding states.  A concrete example 
is provided below in the case of Majorana neutrinos, with the obvious 
generalization to Dirac fermions.

Consider the neutrino sector, which consists of $\nu_e,\nu_\mu,\nu_\tau$ 
in the minimal standard model.  Assume three additional heavy right-handed 
neutrinos $N_{iR}$.  Then the famous canonical seesaw mechanism \cite{seesaw} 
allows all three light neutrinos to acquire mass.  This may be regarded as a 
two-step process.  First, the $3 \times 3$ Majorana mass matrix spanning 
$N_{iR}$ is diagonalized; then the Yukawa coupling matrix linking $\bar 
\nu_{iL}$ to $N_{jR}$ through the usual Higgs scalar doublet $\Phi = (\phi^+,
\phi^0)$ provides a Dirac mass matrix proportional to $v = \langle \phi^0 
\rangle$. The resulting Majorana mass terms of $\nu_i$ are then given by
\begin{equation}
-{\cal L}_m = {v^2 \over 2} \Sigma_j {(\Sigma_i f_{ij} \nu_i)^2 \over M_j} + 
H.c.
\end{equation}
Since all three $\nu_i$'s are contained in the sum over $i$ in the above, 
they can be considered as mass eigenstates (instead of interaction 
eigenstates) just as well.  With this interpretation, the following 
conditions on $f_{ij}$ are obtained:
\begin{equation}
m_i = v^2 \Sigma_j {f_{ij}^2 \over M_j}, ~~~ 0 = \Sigma_j {f_{ij} f_{kj} 
\over M_j}, ~(i \neq k).
\end{equation}
Redefine
\begin{equation}
{f_{ij} v \over \sqrt {M_j}} \equiv \sqrt {m_i} A_{ij},
\end{equation}
then
\begin{equation}
1 = \Sigma_j A_{ij}^2, ~~~ 0 = \Sigma_j A_{ij} A_{kj}, ~(i \neq k),
\end{equation}
which implies that $A_{ij}$ should be an orthogonal matrix.  

The above derivation is actually more general.  All is needed is a Majorana 
mass matrix, then again it can be diagonalized and rewritten in terms of 
the corresponding $\psi_i$'s.  However, the seesaw mechanism is a concrete 
example of how equal ``masses'' may be achieved.  Consider first the $2 
\times 2$ case, i.e.
\begin{equation}
A = \left( \begin{array} {c@{\quad}c} \cos \theta & -\sin \theta \\ 
\sin \theta & \cos \theta \end{array} \right),
\end{equation}
then in particular, $\theta = \pi/4$ may be chosen and each normalized 
mass term of Eq.~(11) becomes the same, i.e.
\begin{equation}
{1 \over 2} (m_1 + m_2).
\end{equation}
A simple realization of this result is to have the original $4 \times 4$ mass 
matrix spanning $\nu_{1,2}$ and $N_{1,2}$ be of the form
\begin{equation}
{\cal M}_{\nu N} = \left[ \begin{array} {c@{\quad}c@{\quad}c@{\quad}c} 
0 & 0 & a & a \\ 0 & 0 & b & -b \\ a & b & M_1 & 0 \\ a & -b & 0 & M_2 
\end{array} \right],
\end{equation}
with $M_1 = M_2$.  The ``mass'' generated by $M_1$ alone is $(a^2 + b^2)/M_1$ 
corresponding to the state $(a \nu_1 + b \nu_2)/\sqrt {a^2 + b^2}$, whereas 
the ``mass'' generated by $M_2$ alone is $(a^2 + b^2)/M_2$ corresponding to 
the state $(a \nu_1 - b \nu_2)/\sqrt {a^2 + b^2}$.  The two ``masses'' are 
thus equal for $M_1=M_2$ but the two states are not orthogonal for $a \neq b$. 
On the other hand, the seesaw reduction of ${\cal M}_{\nu N}$ is
\begin{equation}
{\cal M}_\nu = \left[ \begin{array} {c@{\quad}c} 2a^2/M & 0 \\ 0 & 2b^2/M 
\end{array} \right],
\end{equation}
where $M = M_1 = M_2$.

This shows that the seesaw mechanism may well have started out with the 
generation of two equal masses for two independent linear combinations of 
two neutrinos, which are however not necessarily orthogonal.  It is their 
overlap, i.e. $(m_1-m_2)/(m_1+m_2)$, which determines the mass eigenvalues. 
This scenario with two $N_R$'s is actually very well suited for 
explaining the present data on atmospheric \cite{atm} and solar \cite{sol} 
neutrino oscillations.  [Even though there are only two tree-level neutrino 
masses, the third neutrino is not strictly massless, because it will pick up 
a tiny mass through radiative corrections \cite{bama}.]  Equal ``masses'' 
cannot be implemented in the Majorana case beyond $n=2$ because $A_{ij}$ 
is orthogonal, rather than unitary as in the Dirac case.

In conclusion, it has been shown in this short note that any $n \times n$ 
Dirac fermion mass matrix may be written as the sum of $n$ states of equal 
``mass''.  However, these states are in general not orthogonal.  Thus the 
texture of any such fermion mass matrix may be understood as the nonzero 
overlap among these states.  A concrete example of how this may come about 
is provided by the seesaw mechanism in the $2 \times 2$ case for Majorana 
neutrinos and in the $n \times n$ case for Dirac fermions.\\[5pt]

This work was supported in part by the U.~S.~Department of Energy
under Grant No.~DE-FG03-94ER40837.

\newpage
\bibliographystyle{unsrt}

\end{document}